\documentclass[preprint,aps,showpacs]{revtex4}
\usepackage[latin9]{luainputenc}
\usepackage{float}
\usepackage{amsmath}
\DeclareMathOperator\arctanh{arctanh}
\DeclareMathOperator\arcsinh{arcsinh}

\usepackage{amssymb}
\usepackage{graphicx}
\usepackage{setspace}
\usepackage[unicode=true,pdfusetitle,
 bookmarks=true,bookmarksnumbered=false,bookmarksopen=false,
 breaklinks=false,pdfborder={0 0 1},backref=false,colorlinks=false]
 {hyperref}

\makeatletter

\@ifundefined{textcolor}{}
{%
 \definecolor{BLACK}{gray}{0}
 \definecolor{WHITE}{gray}{1}
 \definecolor{RED}{rgb}{1,0,0}
 \definecolor{GREEN}{rgb}{0,1,0}
 \definecolor{BLUE}{rgb}{0,0,1}
 \definecolor{CYAN}{cmyk}{1,0,0,0}
 \definecolor{MAGENTA}{cmyk}{0,1,0,0}
 \definecolor{YELLOW}{cmyk}{0,0,1,0}
}

\usepackage{amsfonts}
\usepackage{bm}
\usepackage[dvips]{epsfig}
\usepackage[dvips]{graphicx}
\usepackage{float}
\usepackage{dcolumn}
\usepackage{ifpdf}
\usepackage{dcolumn}
\usepackage{multirow}

\newcommand{\sech}{\mathrm{sech}}
\newcommand{\be}{\begin{equation}}
\newcommand{\ee}{\end{equation}}
\newcommand{\bes}{\begin{subequations}}
\newcommand{\ees}{\end{subequations}}
\newcommand{\ben}{\begin{eqnarray}}
\newcommand{\een}{\end{eqnarray}}

\makeatother

\begin{document}
\title{Kink scattering in  hyperbolic models}
 \author{D. Bazeia${^1}$, Adalto R. Gomes$^{2}$, K. Z. Nobrega$^{3}$, Fabiano C. Simas$^{4}$, }
 \email{bazeia@fisica.ufpb.br, argomes.ufma@gmail.com, bzuza1@yahoo.com.br, simasfc@gmail.com}
 \noaffiliation
\affiliation{
$^1$ Departamento de F\'isica, Universidade Federal da Para\'iba, 58051-970, Jo\~ao Pessoa, PB, Brazil\\
$^2$ Departamento de F\'isica, Universidade Federal do Maranh\~ao (UFMA)
Campus Universit\'ario do Bacanga, 65085-580, S\~ao Lu\'\i s, Maranh\~ao, Brazil\\
$^3$ Departamento de Eletro-Eletr\^onica, Instituto Federal de Educa\c c\~ao, Ci\^encia e
Tecnologia do Maranh\~ao (IFMA), Campus Monte Castelo, 65030-005, S\~ao Lu\'is, Maranh\~ao, Brazil\\
$^4$ Centro de Ci\^encias Agr\'arias e Ambientais-CCAA, Universidade Federal do Maranh\~ao
(UFMA), 65500-000, Chapadinha, Maranh\~ao, Brazil
}
\noaffiliation

\begin{abstract}
In this work we study kink-antikink and antikink-kink collisions in hyperbolic models of fourth and sixth order.  We compared the patterns of scattering with known results from polynomial models of the same order. The hyperbolic models considered here tend to the polynomial $\phi^4$ and $\phi^6$ models in the limit of small values of the scalar field. We show that kinks and antikinks that interact hyperbolically with fourth order differ sensibly from those governed by the polynomial $\phi^4$ model.  The increasing of the order of interaction to the sixth order shows that the hyperbolic and polynomial models give intricate structures of scattering that differ only slightly.   The dependence on the order of interaction are related to some characteristics of the models such as the potential of perturbations and the number of vibrational modes.     

\end{abstract}



\maketitle


\section{ Introduction }

Solitary waves are solutions from nonlinear physics characterized by the very special property of localized density energy that can freely propagate without distortion in their form. Topological defects associated with solitary waves are of great interest in several areas of low and high energy physics \cite{vacha,mastu,daupey}.  The simplest solitary wave solution is the   $(1,1)$ dimensional kink and antikink obtained in scalar field theories.

In integrable models, two solitary solutions have the additional property of keep their form even after scattering, acquiring at most a phase shift. These are more properly known as solitons.  In nonintegrable models, on the other hand,  the process of collision can be surprisingly rich. When analyzed as a function of the initial velocity of approximation,  a complicated structure appears \cite{kudry,goodhab}, usually connected with the deformation of the scalar field profile and the emission of scalar radiation.   For large initial velocities a simple inelastic scattering occurs, and after the contact, the kink-antikink pair retreats from each other. On the other hand, for sufficiently small initial velocities, the kink and antikink capture one another and a trapped bion state is formed, that radiates continuously until being completely annihilated. 

An intriguing aspect of the scattering, observed in many nonintegrable models, and in particular in the  very studied $\lambda\phi^4$  model \cite{sugi,moshi,wingate,belkud,camschowin,campeysol,campey,aniolmat,goodhab},
occurs for some windows of intermediate velocities - named two-bounce windows - where the scalar field at the center of mass bounces twice before the pair recedes to infinity. 
Such windows appear in sequence with smaller thickness, accumulating in the border of the one-bounce region.  The same was verified for even higher levels of bounce windows, characterizing a fractal structure \cite{aniolmat}. The formation of two-bounce windows was interpreted phenomenologically in the Ref. \cite{camschowin} as related to the exchange of resonant energy between translational and vibrational modes. The $\phi^6$ model is an exception for this mechanism, despite the absence of vibrational mode in the perturbation of a kink. There it was shown that the resonant scattering appears if we consider the effect of collective modes produced by the pair $\bar K K$ \cite{domerosh}. Another counterexample of the mechanism described in the Ref. \cite{camschowin} is presented in Ref. \cite{sgno}, where two-bounces windows disappear completely despite the presence of vibrational modes. One must remark that the explanation of these intricate processes of scattering, and in particular the formation of two-bounce windows in terms of collective coordinates, was severely criticized in the Ref. \cite{weig1}.

Recently there has been a renewal of the study of  kink and antikink scattering, with several models being subject of investigation. In this line one can cite polynomial models with one \cite{domerosh,dedekecrsa,gakuli1,wei,roman1,belgani1,ganilenliz1,ekagani,fv} and two or more \cite{halromshn,al1,al2,al3} scalar fields, nonpolynomial models \cite{peycam,gankud,sgn,gaaes,bbv,bbv2} and multi-kinks \cite{magasaadmja,maassadm,sadmkev,almasazhdi,gan3}.

The importance of the study of kink scattering can be highlighted with some examples. The context of string landscape has increased the interest in scenarios of bubble collisions. In particular, the collision of relativistic bubbles can be treated as planar walls, described as kink scattering in $(1,1)$ dimensions \cite{gib}.  The realization of kinks in the nonintegrable $\phi^4$ theory was proposed in buckled graphene nanoribbon \cite{graph1, graph2}. Kink-antikink scattering  appear in the study of photogeneration of topological excitations in {\it trans}-polyacetilene. The uncorrelated lattice model of the Su-Schrieffer-Heeger Hamiltonian \cite{ssh} predict that the kink and antikink propagate as independent particles governed by the integrable sine-Gordon equation. A similar behavior was observed in the limit $T=0$ when the quantum mechanical forces between the atoms are computed using hybrid time-dependent density functional theory (TD-DFT) \cite{poly}.  For $T\neq0$, on the other hand, the kink and antikink scatter forming two-bounce windows, a characteristic of non-integrable models \cite{poly}. In this case the realization of the two-bounce is interpreted, when the kink-antikink pair is at short distance, as a competition between the opposite electron-lattice and electron-electron interactions.

The kink and its stability analysis lead to an interesting connection between supersymmetry and quantum mechanics \cite{junker,cookhasuk}. For instance, in a field theory model, the stability analysis of kink structures leads us to a quantum mechanical potential. On the contrary, one can reconstruct a field theory from some characteristic of a quantum mechanical potential
\cite{flosva,boryur,bazbem1,bazlos1}.
In \cite{bfl}, the authors review some results about the deformation procedure in systems with a single real scalar field \cite{bfl,blm,abl}  that give rise to distinct field theories having the same stability potential. In their work, several models with polynomial and hyperbolic interactions were considered.

The presence of scalar field with hyperbolic interactions has appeared some years ago 
in the study of solvable scalar hairy black holes.  One example is the problem of finding a regular configuration of noncharged black hole and the cosmological scalar field \cite{b}. This falsifies Wheeler's conjecture \cite{whee} and evades the scalar ``no-hair" theorems \cite{nh}.  Einstein-scalar gravity with a superscalar potential is obtained after truncating the low energy gauged supergravity limit of string theory \cite{duf,free}. A strategy for obtaining exact solutions of such theories was proposed in the Ref. \cite{c}, that use the symmetry of the equations of motion to give a scale-invariant {\it Ansatz} for the scalar field, leading further to derive the metric and the scalar potential. Within this strategy many classes of exact solutions were obtained, including new scalar hairy black holes with hyperbolic potentials \cite{c}.  

In this work, we consider kink-antikink collisions for two classes of $(1,1)$-dimensional nonintegrable models with $\phi^4$ and $\phi^6$ hyperbolic potentials. In the Sect. II we present our results for kink-antikink scattering for the hyperbolic $\phi^4$ model. We report the suppression of two-bounce windows, despite the presence of vibrational state. In the Sect. III we present our results for the hyperbolic $\phi^6$ model. In the Sect. IV we conclude.


\section{The hyperbolic $\phi^4$ model} \label{secII}


Let us consider the action 
\begin{equation}
S=\int  dtdx \biggl( \frac12 \partial_\mu\phi\partial^\mu\phi - V(\phi) \biggr),
\end{equation}
with the potential of the hyperbolic type \cite{bfl}
\begin{eqnarray}
V(\phi)=\frac14 (1-\sinh^2(\phi))^2.
\label{potential}
\end{eqnarray}
\begin{figure}
\includegraphics[{angle=0,width=8cm}]{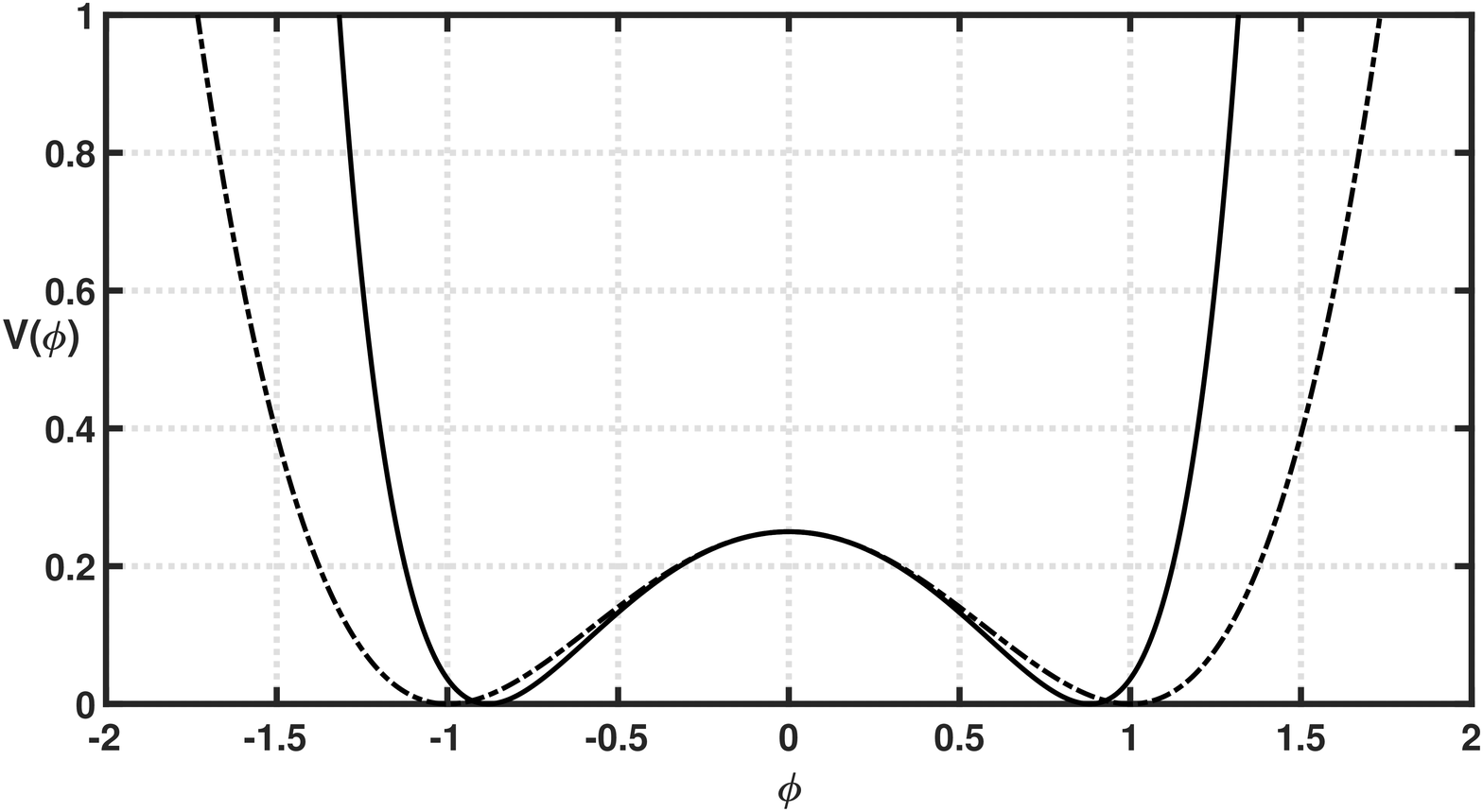}
\includegraphics[{angle=0,width=8cm}]{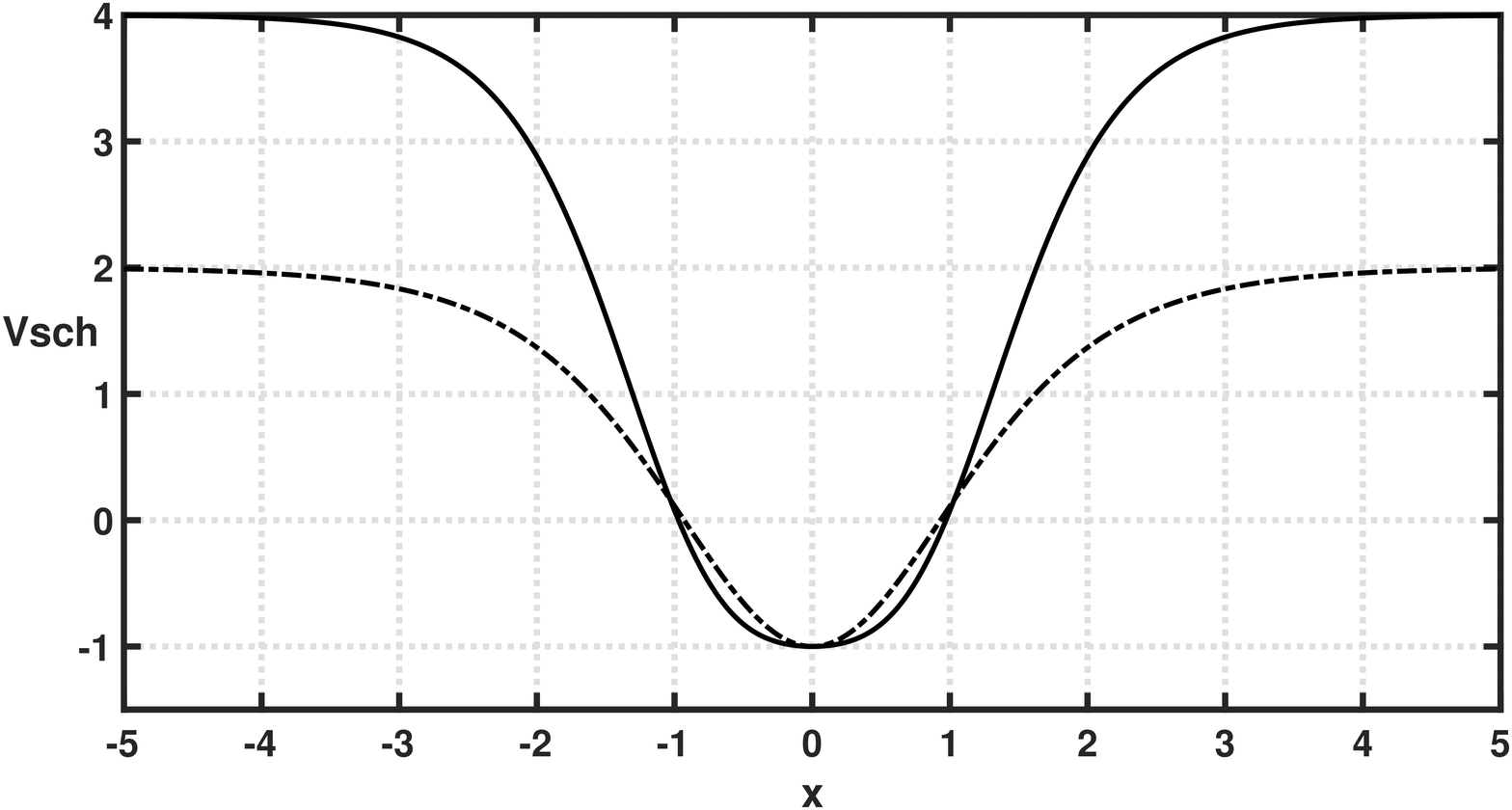}
\caption{The hyperbolic and polynomial $\phi^4$ models: (a) Potential $V(\phi)$ and b) Schrodinger-like potential $V_{sch}(x)$ for perturbations around the kink. Full lines are for the hyperbolic model and traced lines are for the polynomial one.}
\label{fig1}
\end{figure}

The Fig. \ref{fig1}a shows that the potential has two minima in $\phi=\pm\arcsinh(1)$ and one local maximum at the origin. Then we have one
topological sector connecting adjacent minima. We call this model as a hyperbolic $\phi^4$  since, in the limit of small values of $\phi$, it approaches to the usual polynomial $\phi^4$ theory where
\begin{eqnarray}
V(\phi)=\frac14 (1-\phi^2)^2.
\label{potential}
\end{eqnarray}
For comparison we also included in the Fig. \ref{fig1}a the potential for the polynomial $\phi^4$ model.

The equation of motion is
\begin{equation}
\frac{\partial^2\phi}{\partial t^2}-\frac{\partial^2\phi}{\partial x^2}+\frac{dV}{d\phi}=0.
\end{equation}
Static kink solution is given by \cite{bfl}
\begin{eqnarray}
  \phi(x) = \arctanh\Bigg(\frac{1}{\sqrt{2}} \tanh(x) \Bigg).
    \label{solution2}
\end{eqnarray}
Corresponding antikink solution is given by $\phi_{\bar K}(x)=\phi_K(-x)$. The energy density is given by
\begin{eqnarray}
   \rho(x)=\frac{2\sech^4(x)}{(2-\tanh^2(x))^2}.
\end{eqnarray}
Perturbing linearly the scalar field around one kink solution $\phi_K(x)$ as $\phi(x,t)=\phi_K(x)+\eta(x)\cos(\omega t)$ we get a Schr\"odinger-like equation
\be
-\frac{\partial^2\eta}{\partial x^2}+ V_{sch}\,\eta=\omega^2\eta.
\ee
with the Schr\"odinger-like potential
\be
V_{sch}(x)=\frac{d^2 V}{d\phi^2} = \frac{2 \big(\tanh^4(x)+3\tanh^2(x)-2  \big)}{\big( \tanh^2(x)-2\big)^2}.
\ee
This stability potential for the kink is presented in the Fig. \ref{fig1}b. We have the same potential $V_{sch}(x)$ for the antikink $\phi_{\bar K}(x)$. This potential is broader and shallower than the polynomial $\phi^4$ model, shown in the same figure. This favors the occurrence of more vibrational modes. Indeed, it has three eigenvalues, corresponding to the zero mode (translational mode) and two vibrational modes ($\omega_1=1.4325$ and $\omega_2=1.9113$). Compare with the two eigenvalues for the polynomial $\phi^4$ model \cite{sugi} ($\omega_0=0$ and $\omega_1=\sqrt{3/2}\simeq 1.2247$). From the structure of the potentials we note that the bound states must be centered at $x=0$, corresponding to the minimum of $V_{sch}$. As we remarked in the previous section, the vibrational modes are important for the understanding of the intricate collision process. 

Now we present our main results of kink-antikink scattering. The results for antikink-kink scattering are identical. We solved the equation of motion with a pseudospectral on a grid with $2048$ nodes with periodic boundary conditions. We fixed $x=\pm x_0$ with $x_0 = 12$ for the initial symmetric position of the pair and set the grid boundary at $x=\pm x_{max}$ with $x_{max}=400$. A sympletic method with the Dirichlet condition imposed at the boundaries was also applied to double check our numerical results. In this case
 it was used a $4^{th}$ order finite-difference method with spatial step $\delta x=0.08$ and a $6^{th}$ order symplectic integrator with time step $\delta t=0.04$.
\begin{figure}
\includegraphics[{angle=0,width=8cm}]{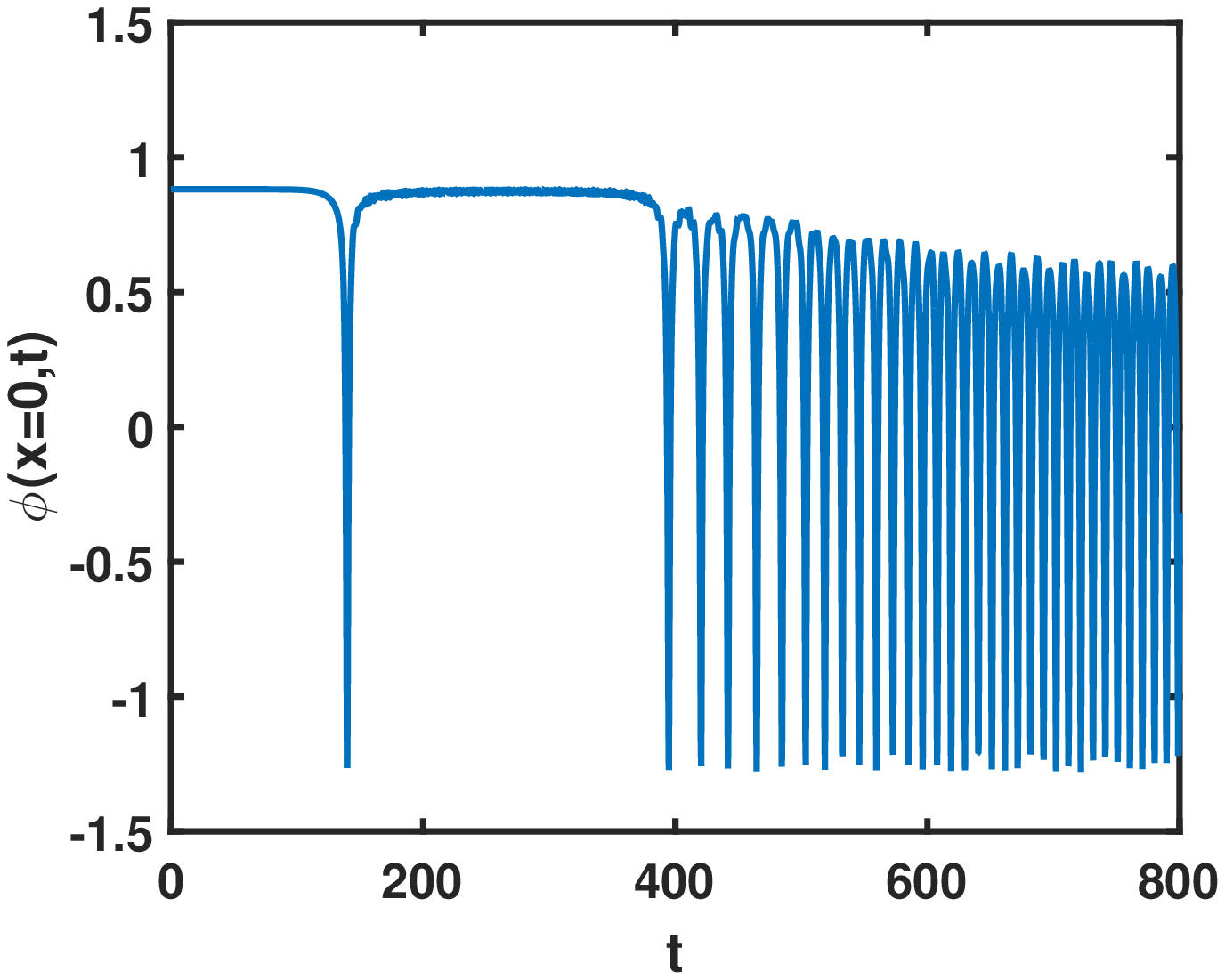} 
\includegraphics[{angle=0,width=8cm}]{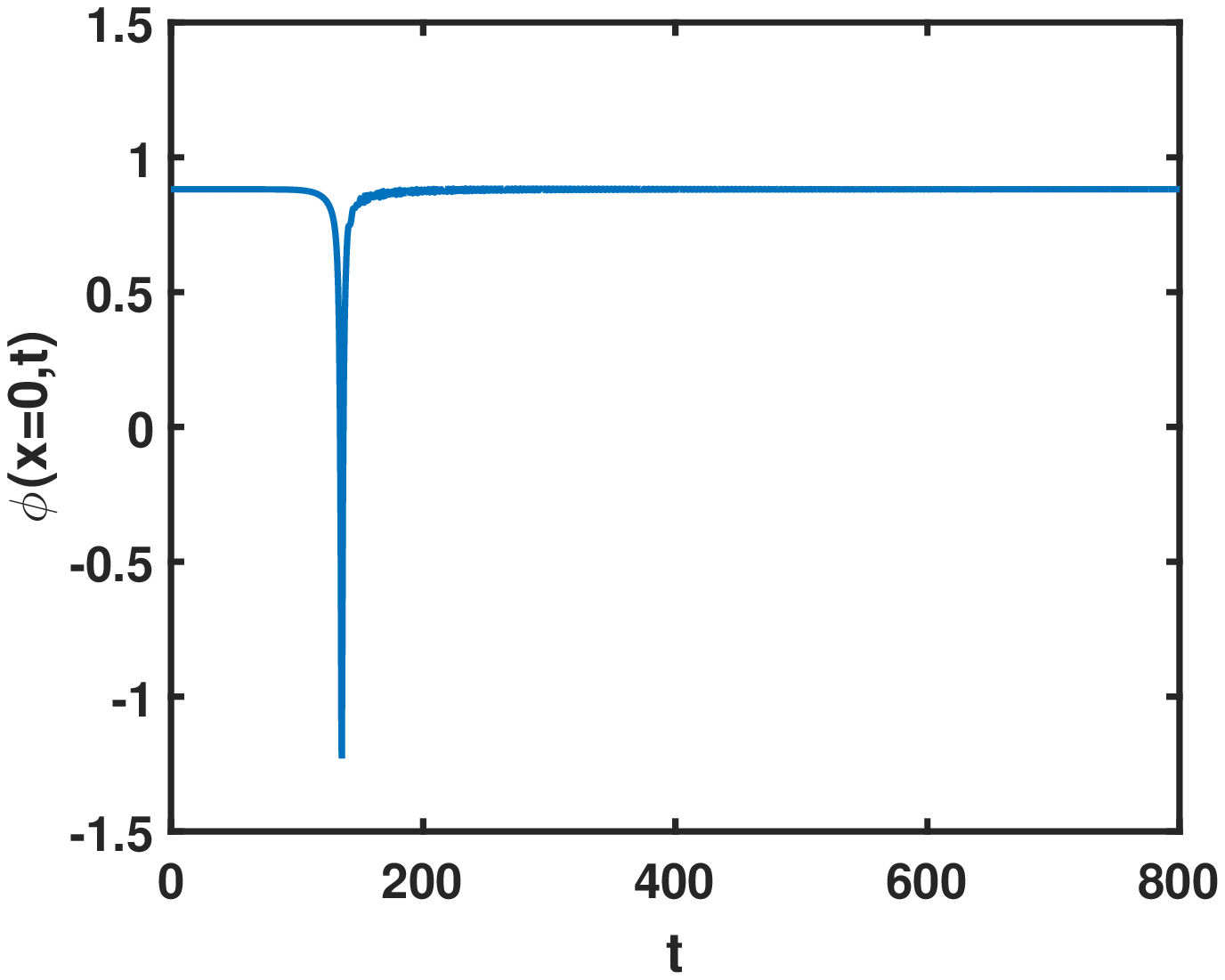} 
\caption{Kink-antikink collisions for the hyperbolic $\phi^4$ model: scalar field at the center of mass $\phi(x=0,t)$ versus time for (a) $v=0.058$ and (b)  $v=0.06$.}
\label{fig2}
\end{figure}
For solving the equation of motion for kink-antikink scattering we used the following initial conditions
\begin{eqnarray}
\phi(x,0)&=&\phi_K(x+x_0,v,0)-\phi_{K}(x-x_0,-v,0)+\arcsinh(1),\\
\dot\phi(x,0)&=&\dot\phi_K(x+x_0,v,0)-\dot\phi_{K}(x-x_0,-v,0),
\end{eqnarray}
where $\phi_K(x+x_0,v,t)$ means a boost solution for kink. 

For $v<v_c$ with $v_c = 0.0587$, bion states are achieved, where the scalar field at the center of mass $\phi(0,t)$ changes after the scattering from the initial value $\phi=\arcsinh(1)$ to erratic oscillations around the adjacent vacuum $\phi=-\arcsinh(1)$. An example of this is shown in the Fig. \ref{fig2}a. After long time emitting scalar radiation, the kink-antikink pair annihilates and the scalar field goes to
the vacuum $\phi=-\arcsinh(1)$. For $v>v_c$ the output is an inelastic scattering between the pair. In this case, $\phi(0,t)$ shows
one-bounce around the vacuum $\phi=\arcsinh(1)$ - see an example in the Fig. \ref{fig2}b.
\begin{figure}
		\includegraphics[{angle=0,width=12cm}]{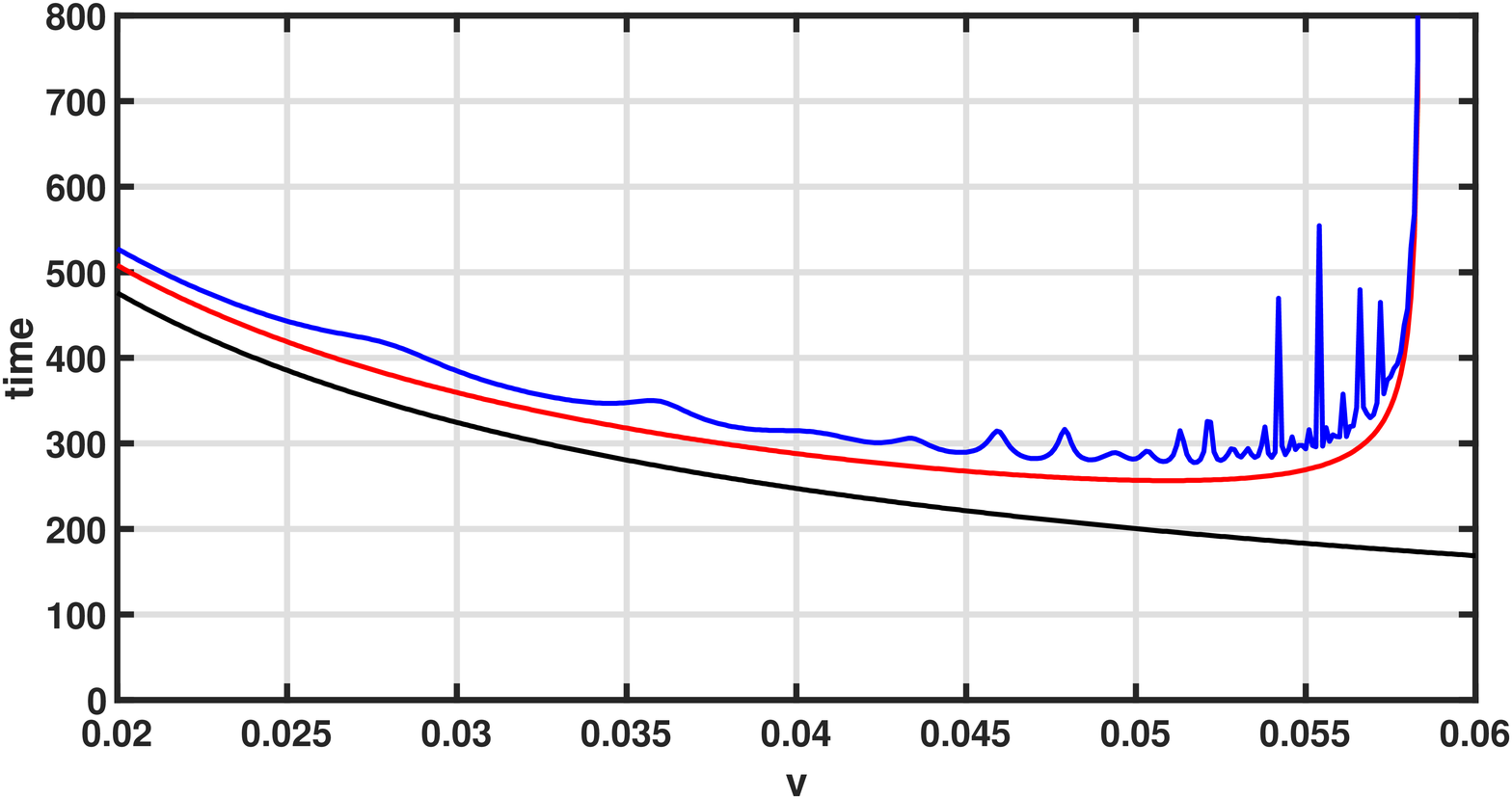}\\
	\includegraphics[{angle=0,width=12cm}]{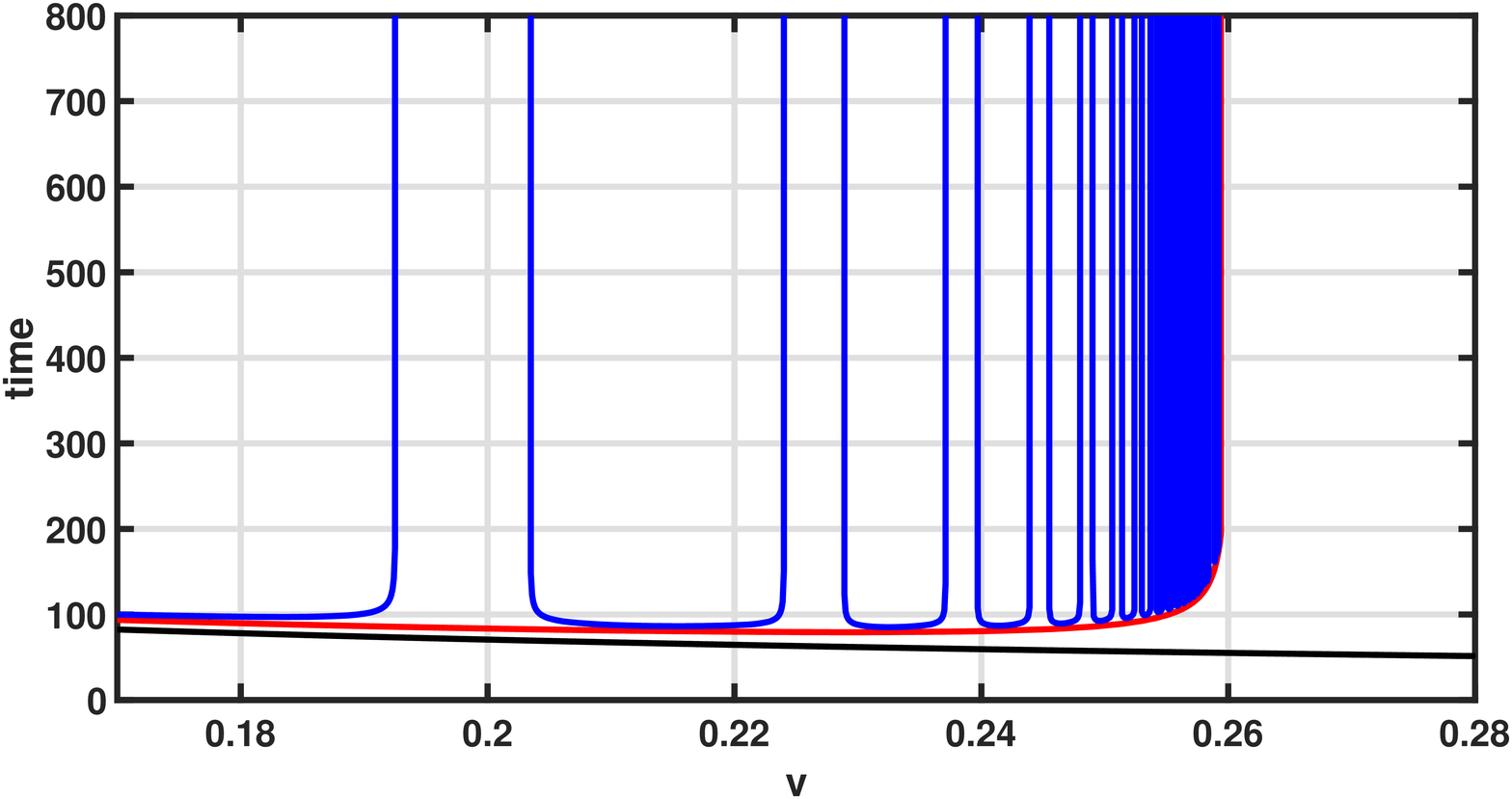}
\caption{Kink-antikink collisions fot the $\phi^4$ model: the times to the first (black), second (red) and third (blue) bounces for antikink-kink collisions as a function of initial velocity for the (a) (upper) hyperbolic $\phi^4$ model  and (b) (lower) polynomial $\phi^4$ model (see the Refs. \cite{aniolmat,goodhab}).}
	\label{phi4}
\end{figure}

The structure of scattering fot the hyperbolic $\phi^4$ model is depicted in the Fig. \ref{phi4}a. We found no two-bounce windows. This is related to the presence of more than one vibrational mode. Indeed, during the scattering, the energy of the translational mode can be transferred to the two vibrational modes. This makes more difficult the realization of the mechanism of resonant energy exchange between the translational and one vibrational mode. As a comparison, we show in the Fig. \ref{phi4}b the known results for the polynomial $\phi^4$ model \cite{sugi,moshi,wingate,belkud,camschowin,campeysol,campey,aniolmat,goodhab}, where the presence of two-bounce windows are related to the existence of only one vibrational mode. Note that for the polynomial model, the critical velocity $v_{crit}=0.259$ is larger than the observed $v_c = 0.0587$ for the hiperbolic model. Note also that for $v\lesssim v_{crit}$ the time interval between the first to the second bounces can be quite large for the hyperbolic $\phi^4$ model (Fig. \ref{phi4}a) in comparison to the observed for the polynomial $\phi^4$ model (Fig. \ref{phi4}b). This corresponds to a behavior of false one-bounce, followed by a bion state, as shown in the Fig. \ref{fig2}a.

\section{The hyperbolic $\phi^6$ model} \label{secIII}

Now we consider the potential \cite{bfl}
\begin{eqnarray}
V(\phi)=\frac12 \tanh^2(\phi) (1-\sinh^2(\phi))^2.
\label{potentialphi6}
\end{eqnarray}
\begin{figure}
\includegraphics[{angle=0,width=8cm}]{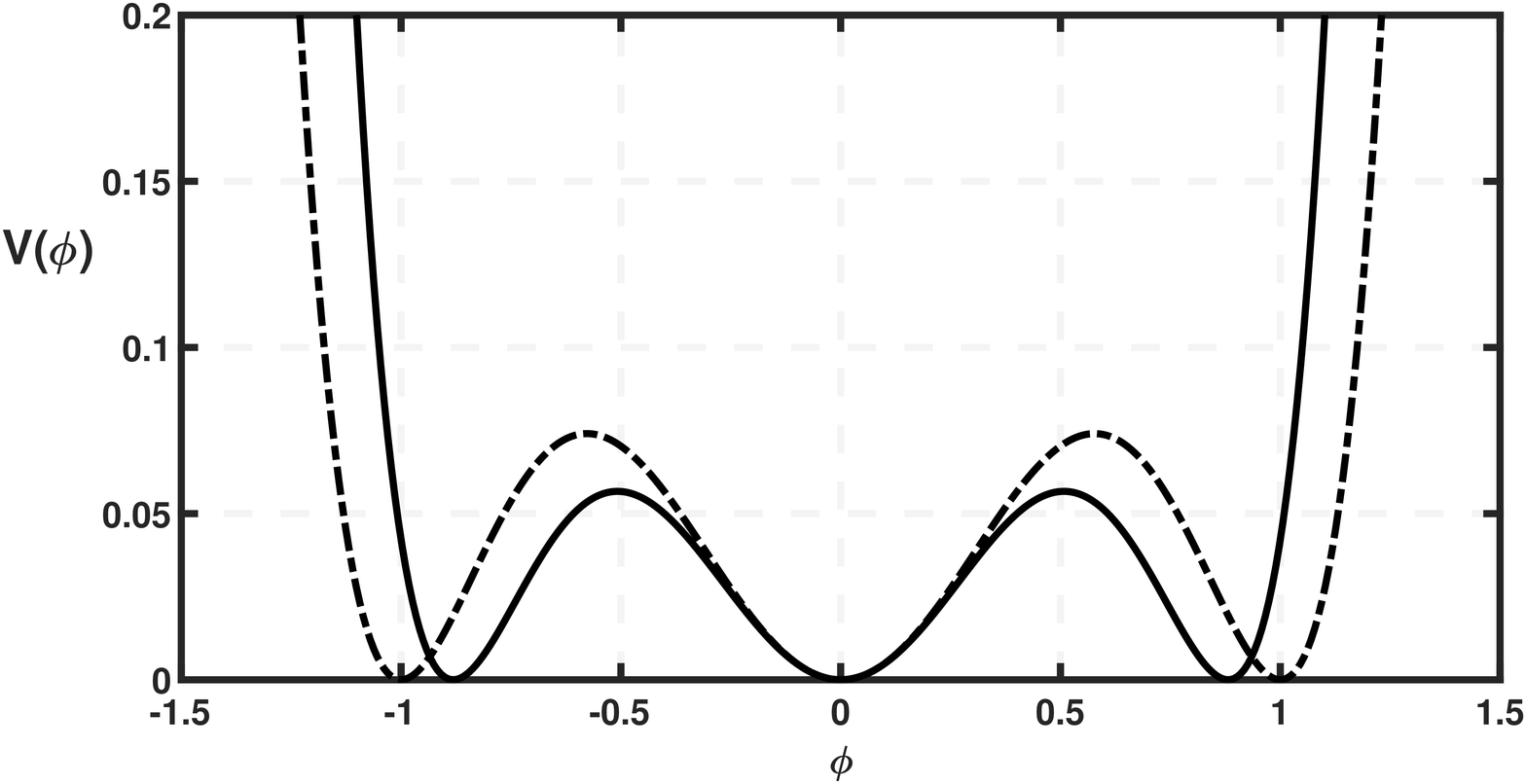}
\includegraphics[{angle=0,width=8cm}]{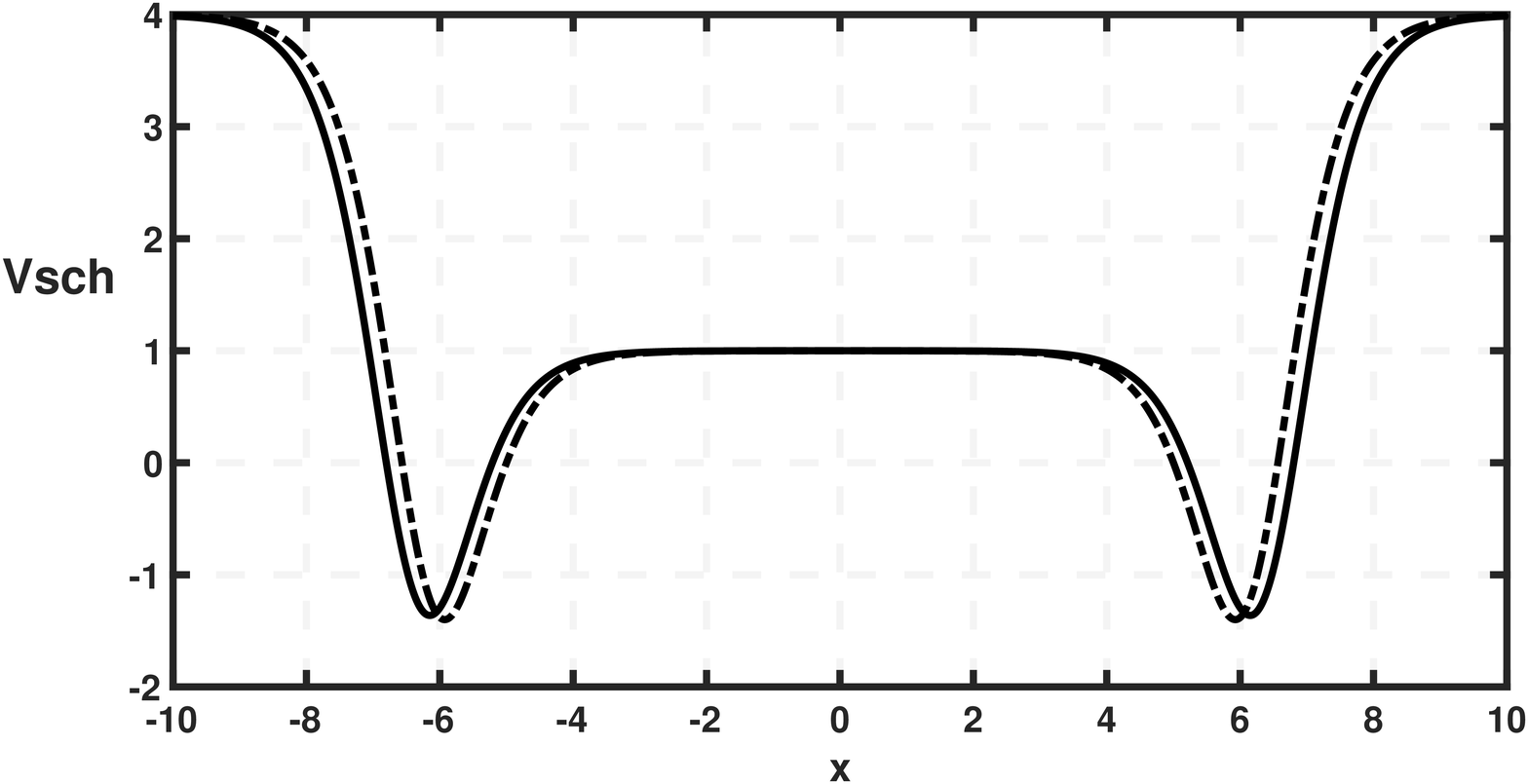}
\caption{The hyperbolic and polynomial $\phi^6$ models: (a) Potential $V(\phi)$ and (b) Schrodinger-like potential $V_{sch}(x)$ for perturbations around the antikink-kink pair. Full lines are for the hyperbolic model and traced lines are for the polynomial one.}
\label{fig1phi6}
\end{figure}
Fig. \ref{fig1phi6}a shows that the potential has three minima at $\phi=0, \pm \arcsinh(1)$ and two local maximum at $\phi=\pm 1/2$. Then we have two
topological sectors connecting adjacent minima. We call this model as a hyperbolic $\phi^6$  since, in the limit of small values of $\phi$, it approaches a polynomial $\phi^6$ theory studied in the Ref. \cite{domerosh}, where
\begin{eqnarray}
V(\phi)=\frac12 \phi^2(1-\phi^2)^2.
\label{potential}
\end{eqnarray}
In the sector $(0,\arcsinh(1))$ we have the following static antikink ($\phi_{\bar K}$) and kink ($\phi_K$) solutions:
\begin{eqnarray}
  \phi_{\bar K} = \arctanh\Bigg[\frac{1}{\sqrt2} \sin \bigg(\frac12 \arccos(\tanh(x)) \bigg)  \Bigg],
\label{solution1phi6}
\end{eqnarray}
\begin{eqnarray}
  \phi_K=  \arctanh\Bigg[\frac{1}{\sqrt2} \cos \bigg(\frac12 \arccos(\tanh(x)) \bigg)  \Bigg].
   \label{solution2phi6}
\end{eqnarray}
For the sector $(0,-\arcsinh(1))$ we have also a kink ($\phi_{\bar K}$) and an antikink ($-\phi_K$) solution.  Without loosing generality, we will work in the topological sector connecting the vacua $\phi=0$ and $\phi=\arcsinh(1)$.

\begin{figure}
	\includegraphics[{angle=0,width=10cm}]{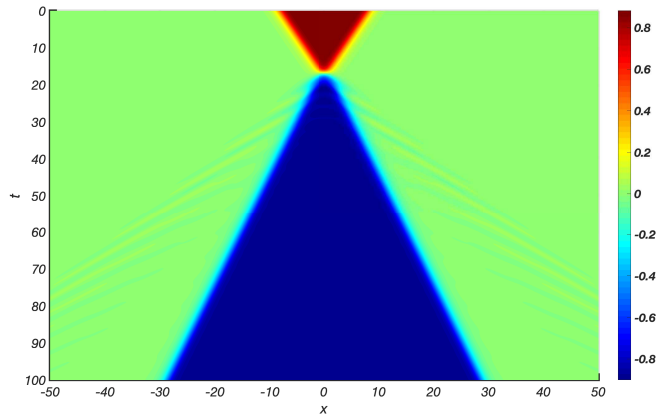}
	\caption{The hyperbolic $\phi^6$ model: kink-antikink with $v=0.457$, showing a one-bounce collision with exchange of topological sector.}
	\label{fig0phi6}
\end{figure}

Stability analysis of this model shows that the Schr\"odinger-like potential for the kink or antikink has no additional bound state besides the zero mode. This could in principle forbid the occurrence of two-bounce windows in a scattering process. However, following the known strategy for the polynomial $\phi^6$ model \cite{domerosh}, we consider the perturbations around the pairs kink-antikink and antikink-kink. We found Schr\"odinger-like potentials that depend on the initial configuration of the pair.

The kink-antikink configuration is given by  $\phi_K(x+x0)+\phi_{\bar K}(x-x0)-\arcsinh(1)$. As a result, the potential of perturbations has a central barrier. This configuration does not present bound states, and two-bounce windows are absent. The scattering process shows bion state for $v<v_{crit}=0.279$, whereas for $v>v_{crit}$ we have one-bounce collisions with an exchange of topological sector, as shown in the Fig.\ref{fig0phi6}. 

The antikink-kink configuration is given by  $\phi_{\bar K}(x+x0)+\phi_K(x-x0)$. We see in the Fig. \ref{fig1phi6}b that the potential of perturbations has the form of a well around $x=0$. This allows the appearance of bound states. This potential is similar to the obtained for the polynomial $\phi^6$ model 
\cite{domerosh}. The potentials of perturbations for the antikink-kink give the following vibrational modes:  $\omega=1.034$, $1.25$, $1.57$ and $1.88$ for the hyperbolic $\phi^6$, and  $\omega=1.045$, $1.28$, $1.61$ and $1.92$ for the polynomial $\phi^6$.

\begin{figure}
\includegraphics[{angle=0,width=8.5cm}]{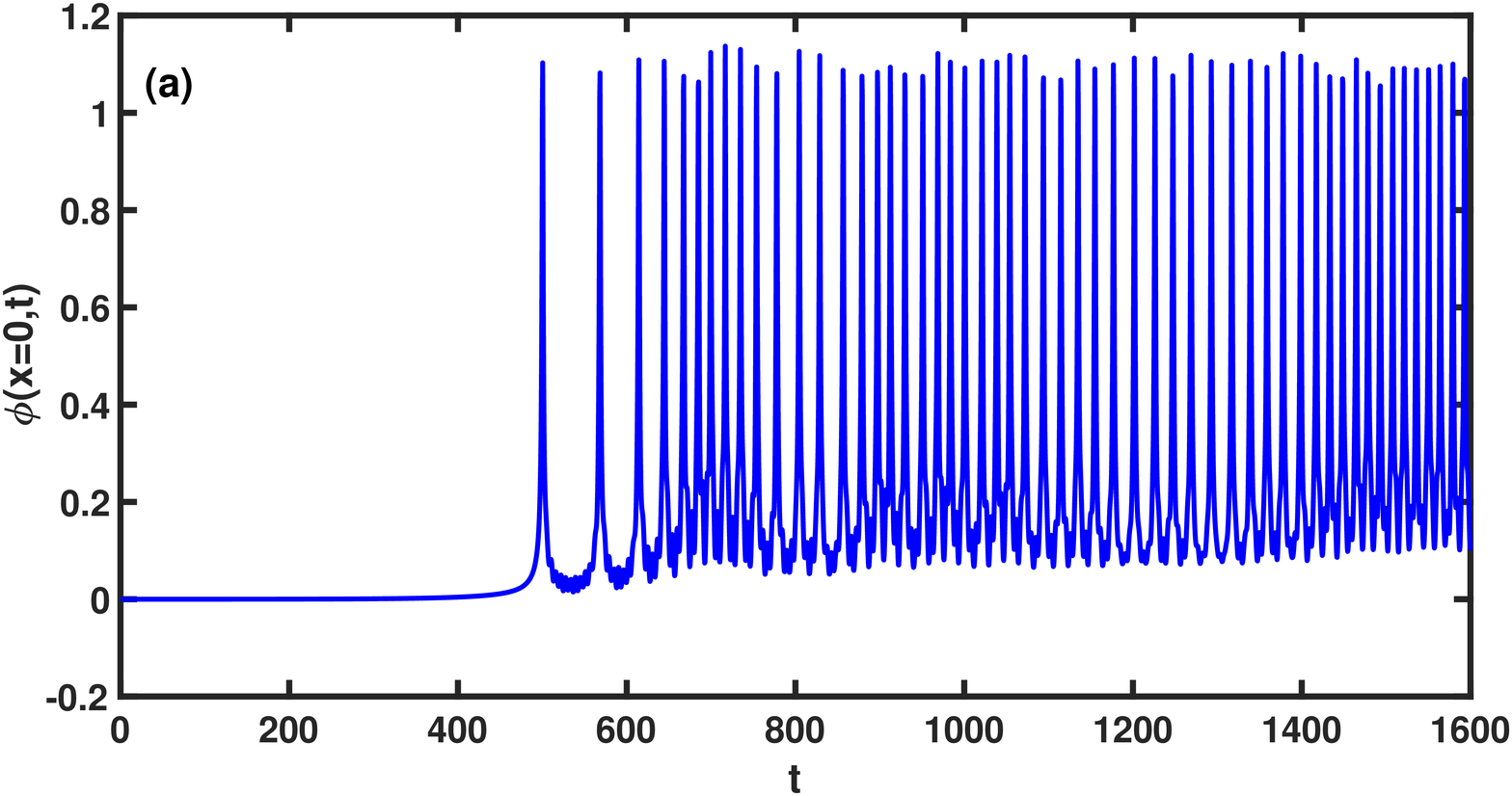}
\includegraphics[{angle=0,width=8.5cm}]{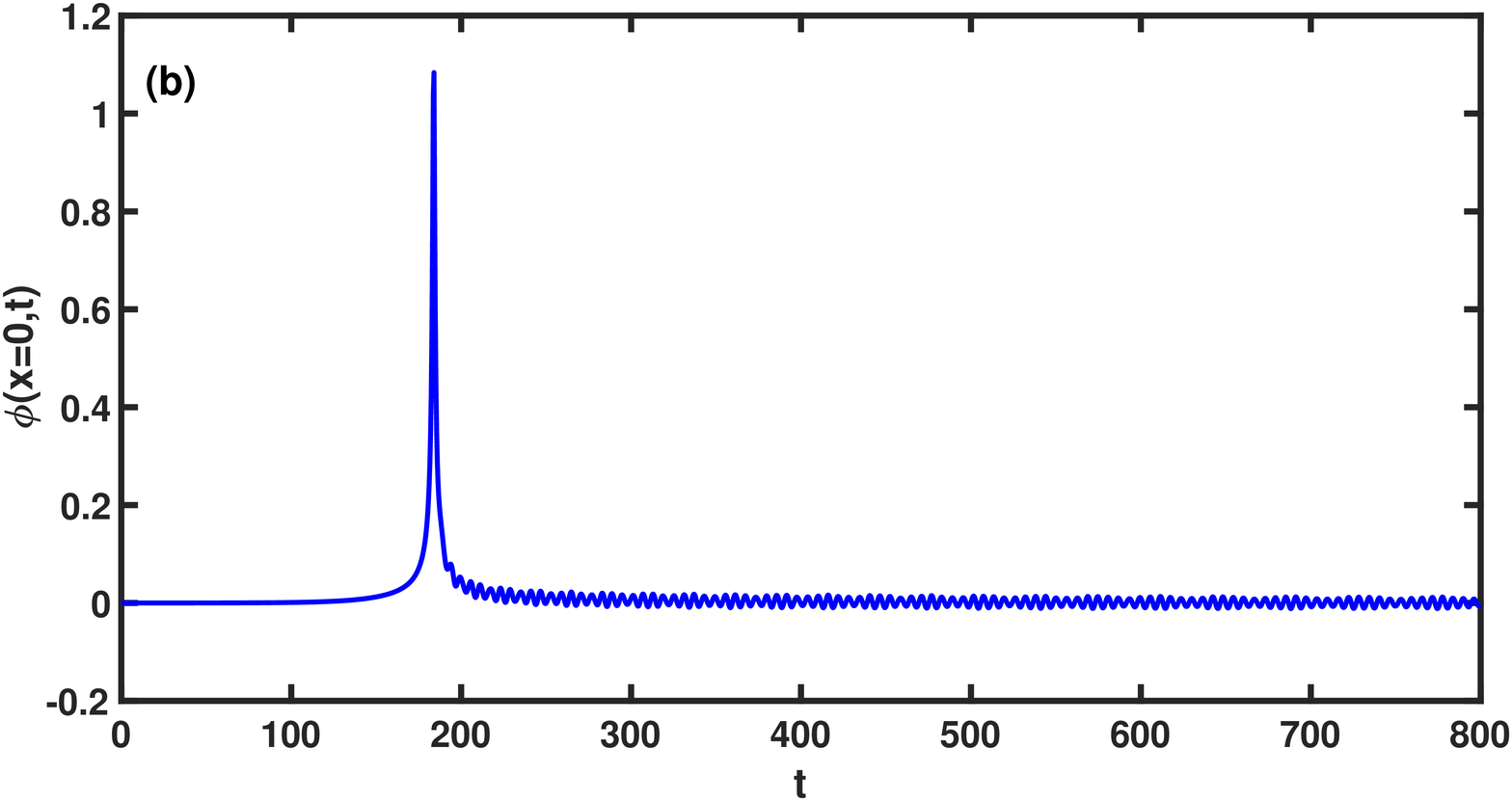}\\
\includegraphics[{angle=0,width=8.5cm}]{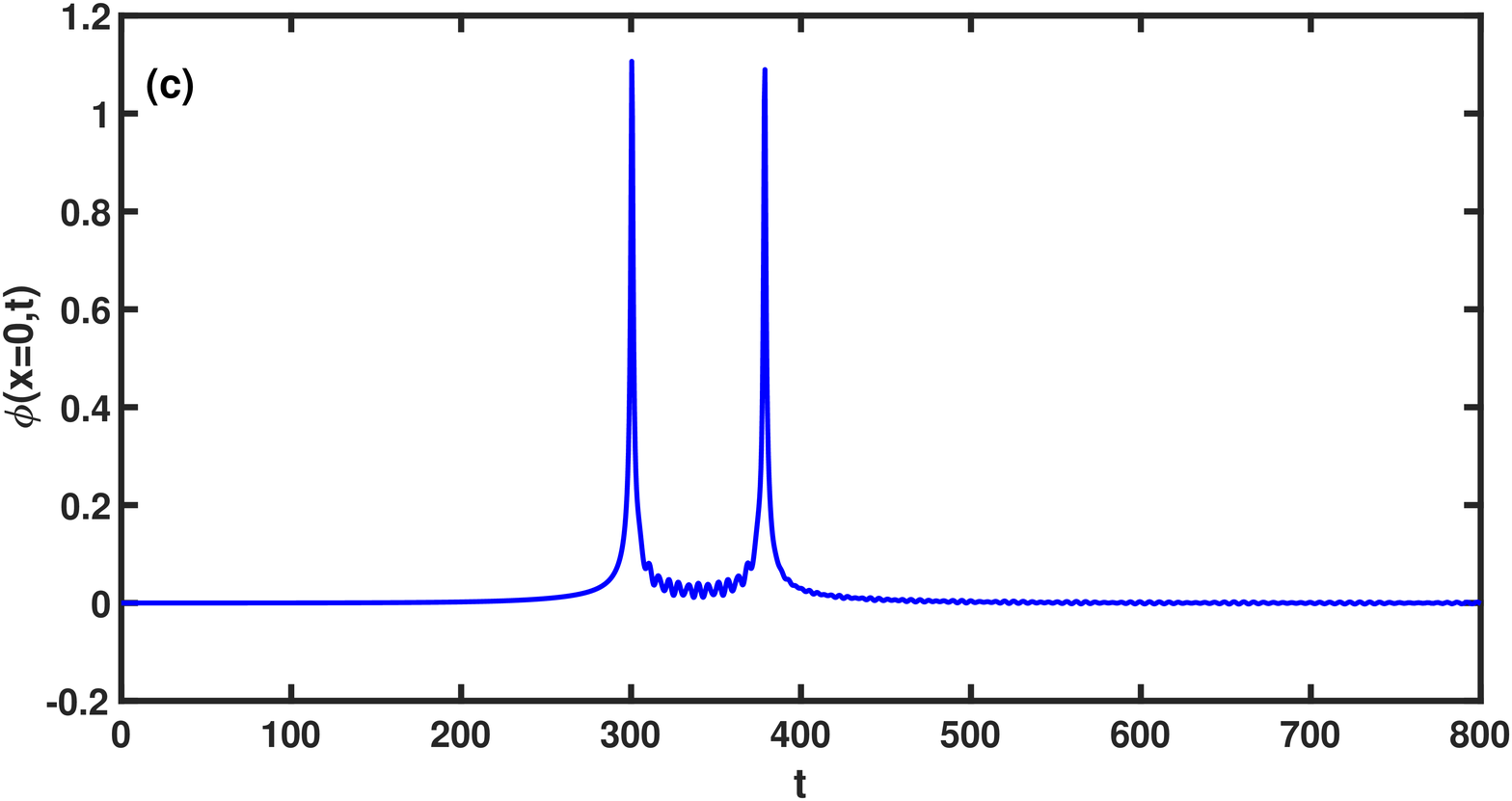}
\includegraphics[{angle=0,width=8.5cm}]{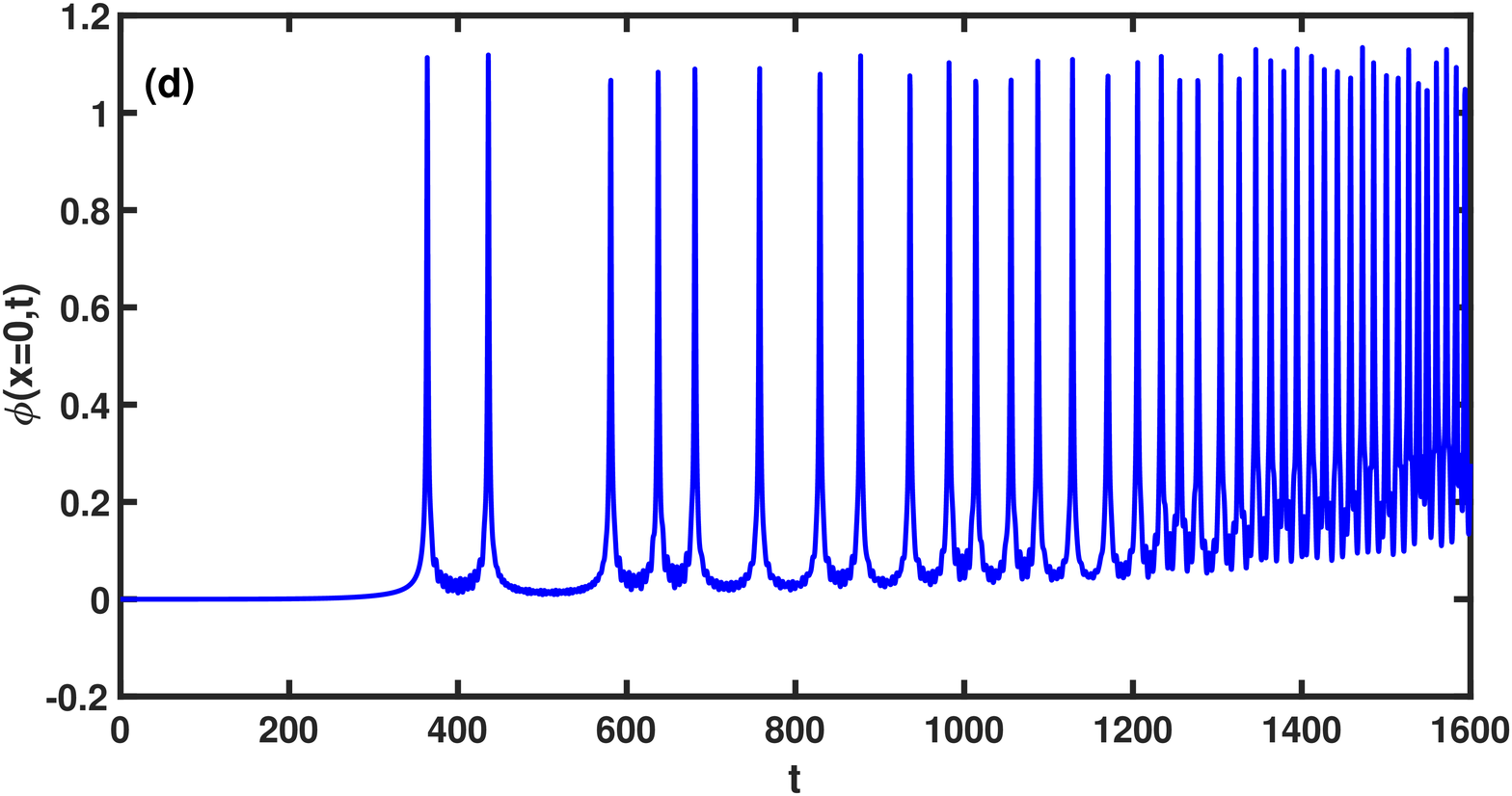}\\ 
\caption{Antikink-kink collisions for the hyperbolic $\phi^6$ model: scalar field at the center of mass $\phi(x=0,t)$ versus time for (a) $v=0.0154$, (b) $v=0.0482$, (c) $v=0.0276$ and (d) $v=0.0222$. To make easier the visualization, we used larger intervals of time in the Figs. (a) and (d).}
\label{fig2phi6}
\end{figure}

Now we present our main results of antikink-kink scattering. The numerical method was the same used for the hyperbolic $\phi^4$ model. We used the following initial conditions
\begin{eqnarray}
\phi(x,0)&=&\phi_{\bar K}(x+x_0,v,0)+\phi_K(x-x_0,-v,0),\\
\dot\phi(x,0)&=&\dot\phi_{\bar K}(x+x_0,v,0)+\dot\phi_K(x-x_0,-v,0),
\end{eqnarray}
For $v<v_c$ with $v_c = 0.0482$, we have bion states, and the scalar field at the center of mass  changes after the scattering from the initial value $\phi=0$ to erratic oscillations, as shown in Fig. \ref{fig2phi6}a.  After long time emitting scalar radiation, the antikink-kink pair annihilates and the scalar field goes to
the vacuum $\phi=+\arcsinh(1)$. 
For $v>v_c$ the output is an inelastic scattering between the pair, with  $\phi(0,t)$ showing
one-bounce around the vacuum $\phi=0$ - see Fig. \ref{fig2phi6}b. 
Also, for some windows in velocities $v \lesssim v_c$, $\phi(0,t)$ presents two-bounce around the vacuum $\phi=0$, as shown in Fig. \ref{fig2phi6}c. In the Fig. \ref{fig2phi6}d we see the appearance of a false two-bounce windows. One can see that after the second bounce $\phi(x=0,t)$ stay for a quite long time around the vacuum $\phi=0$,  with the field oscillating further in a bion pattern and tending in the long run to the vacuum $\phi=+\arcsinh(1)$. 

\begin{figure}
	\includegraphics[{angle=0,width=12cm}]{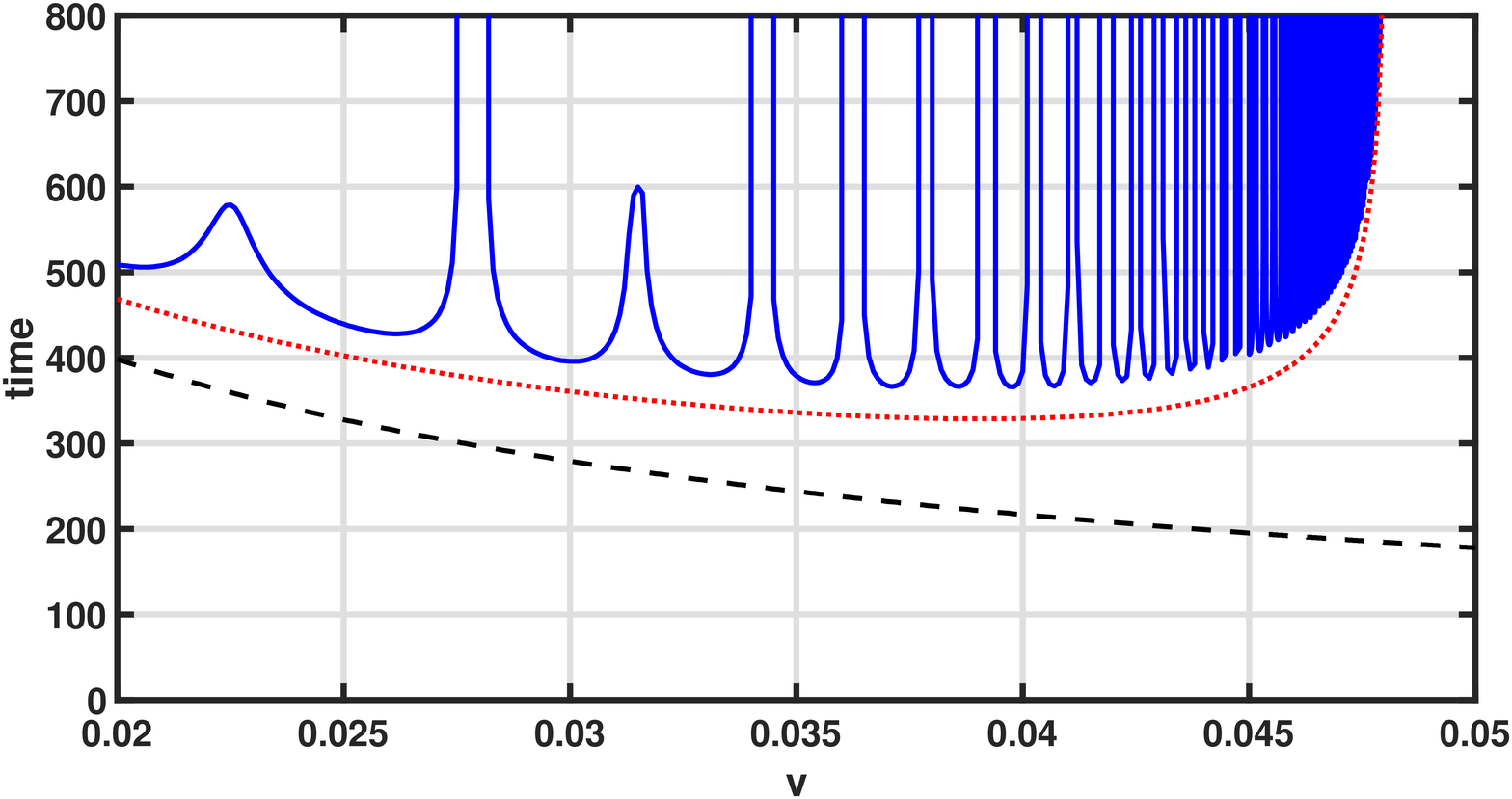} 
   \includegraphics[{angle=0,width=12cm}]{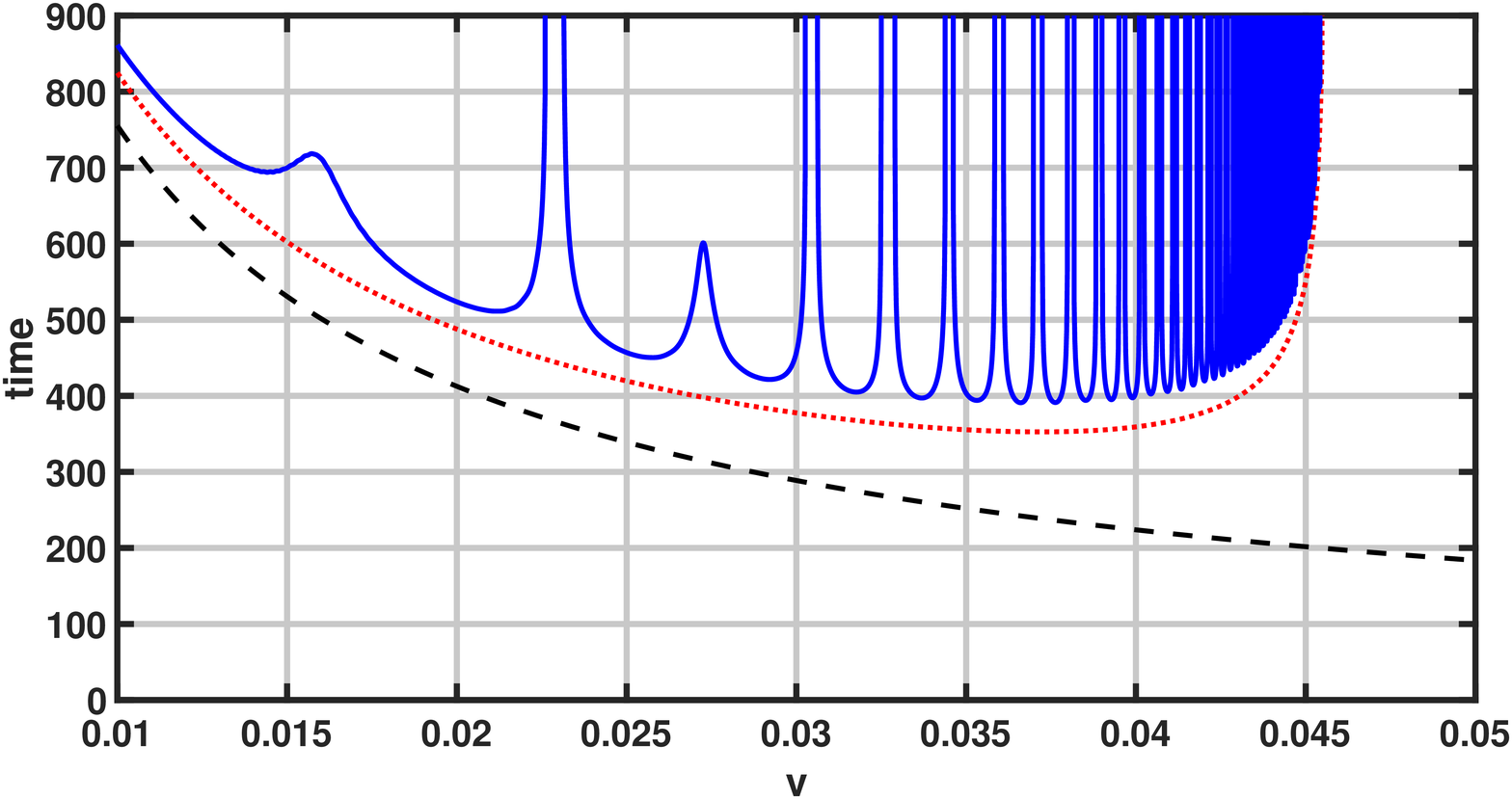}\\
	
\caption{The times to the first (black), second (red) and third (blue) antikink-kink collisions as a function of initial velocity for the (a) $\phi^6$ hyperbolic model and (b) $\phi^6$ polynomial model (see the Ref. \cite{domerosh}).}
\label{times-fig6}
\end{figure}
The Fig. \ref{times-fig6}a shows the time interval $t$ for the three first bounces as a function of the initial velocity $v$. The two-bounce windows are visible at the intervals where the time for the third antikink-kink collisions diverge. The peaks in blue corresponds to false two-bounce windows. In particular, the case presented in the Fig. \ref{fig2phi6}d corresponds to the first peak of the Fig. \ref{times-fig6}a.  We see also that the thickness of the true two-bounce windows is reduced when the velocity approaches $v_c$ from bellow.  For comparison, we present in the Fig. \ref{times-fig6}b the known results from the scattering of the polynomial $\phi^6$ model \cite{domerosh}. Note the similarity of the results from both models. 

The order of each two-bounce window is related to the number $m$ of oscillations of $\phi(0,t)$ between the bounces. For instance, in the Fig. \ref{fig2phi6}d, a false two-bounce windows, we have $m=12$, corresponding to the first peak of the Fig. \ref{times-fig6}a . For the  Fig. \ref{fig2phi6}c we have  $m=13$, corresponding to the first two-bounce window of the Fig. \ref{times-fig6}a. This also happens with the polynomial $\phi^6$ model, where the expected 11 first windows do not appear. This can be related to the presence of two or more vibrational states, as reported  for another model in the Ref. \cite{sgno}. We can compare with the polynomial $\phi^4$ model that has only one vibrational state. There the order of two-bounce windows corresponds exactly to the number $m$  of oscillations, starting with $m=1$ \cite{aniolmat}.

\section { Conclusions  }

We noted that the hyperbolic and polynomial $\phi^4$ models differ sensibly in the potential $V_{sch}$ of perturbations. Comparing both models, the potential $V_{sch}$ for the hyperbolic kink is shallower and thicker, favoring the occurrence of more vibrational states - two for the hyperbolic $\phi^4$ ($\omega=1.4325, 1.9113$)  in comparison to only one ($\omega = 1.2247$) for the polynomial $\phi^4$. This has strong effects in the structure of scattering. Indeed, the polynomial $\phi^4$ model has a structure of two-bounce windows, whereas for the hyperbolic  $\phi^4$ model the two-bounce windows are totally suppressed. The suppression is a kind of destructive interference between the two vibrational modes that forbids the realization of the resonance mechanism of transferring of energy from the translational mode to the vibrational mode. Such effect of total suppression was already described in other models \cite{sgno}. The transition of a region of bion to one-bounce scattering above a critical velocity was observed in both models. However, the critical velocities differ sensibly: $v_{crit}=0.0587$ for the hyperbolic $\phi^4$ and $v_{crit}=0.259$ for the polynomial $\phi^4$. This shows that the kink-antikink pair with a fourth-order hyperbolic interaction must be initially at a very low velocity to arrive a bion state and further annihilate into scalar radiation. In the majority of scenarios the kink-antikink pair scatter inelastically.

For the sixth-order we have a different situation. There  both models, polynomial and hyperbolic, are only slightly distinguishable. Following the procedure described in the Ref. \cite{domerosh}, one must analyze the potential of perturbations of the pair. This results that only antikink-kink result in a Schr\"odinger-like potential with bound states besides the zero mode. For this case it was found that the frequences of the four vibrational states do not differ sensibly: $\omega=1.034$, $1.25$, $1.57$ and $1.88$ for the hyperbolic $\phi^6$ and  $\omega=1.045$, $1.28$, $1.61$ and $1.92$ for the polynomial $\phi^6$. The number of vibrational states is connected to presence of two-bounce windows, false two-bounce windows and the lack of some expected windows. The critial velocities also does not differ sensibly: $v_{crit}=0.0457$ for the hyperbolic $\phi^6$ and $v_{crit}=0.0482$ for the polynomial $\phi^6$.


\section{Acknowledgements}
A.R.G, F.C.S. and K.Z.N. thank FAPEMA - Funda\c c\~ao de Amparo \`a Pesquisa e ao Desenvolvimento do Maranh\~ao through grants PRONEX $01452/14$, PRONEM $01852/14$, Universal-$01061/17$, Universal-$01332/17$, Universal-$01191/16$, Universal-$01441/18$. A.R.G thanks CNPq (brazilian agency) through grants $309842/2015$-8, $311501/2018$-4, $437923/2018$-5 for financial support. This study was financed in part by the Coordena\c c\~ao de Aperfei\c coamento 
de Pessoal de N\' ivel Superior - Brasil (CAPES) - Finance Code 001.
D.B. thanks CNPq for financial support. 


\begin{thebibliography}{99}

\bibitem{vacha} T. Vachaspati, Kinks and Domain Walls: An Introduction to Classical and Quantum Solitons (Cambridge University Press, Cambridge, England, 2006).


\bibitem{mastu} N. Manton and P. Sutcliffe, Topological Solitons (Cambridge University Press, Cambridge, England, 2004).

\bibitem{daupey} T. Dauxois and M. Peyrard, Physics of Solitons, Cambridge University Press, Cambridge (2006).

\bibitem{kudry} T I Belova, A E Kudryavtsev, Solitons and their interactions in classical field theory, Physics-Uspekhi 40 (4) 359 (1997).

\bibitem{goodhab} R.H. Goodman and R. Haberman, {\it Kink-Antikink Collisions in the $\phi^4$ Equation: The n-Bounce Resonance and the Separatrix Map}, SIAM J. Appl. Dyn. Syst. 4 (2005) 1195.

\bibitem{sugi} T. Sugiyama, {\it Kink-Antikink Collisions in the Two-Dimensional $\phi^4$ Model}, Prog. Theor. Phys. 61 (1979) 1550.

\bibitem{moshi} M. Moshir, {\it Soliton-antisoliton scattering and capture in $\phi^4$ theory}, Nucl. Phys. B 185 (1981) 318.

\bibitem{wingate} C.A. Wingate, {\it Numerical Search for a $\phi^4$ Breather Mode}, SIAM J. Appl. Math. 43 (1983) 120.

\bibitem{belkud} T.I. Belova and A.E. Kudryavtsev, {\it Quasi-periodic orbits in the scalar classical $\lambda \phi^4$ field theory}, Physica D 32 (1988) 18.

\bibitem{camschowin} D.K. Campbell, J.S. Schonfeld and C.A. Wingate, {\it Resonance Structure in Kink-antikink interactions in $\phi^4$ theory}, Physica D 9 (1983) 1.

\bibitem{campeysol} D.K. Campbell, M. Peyrard and P. Soldano, {\it Kink-antikink interactions in the double sine-Gordon equation}, Physica D 19 (1986) 165.

\bibitem{campey} D.K. Campbell and M. Peyrard, {\it Solitary wave collisions revisited}, Physica D 18 (1986) 47.

\bibitem{aniolmat} P. Anninos, S. Oliveira and R.A. Matzner, {\it Fractal structure in the scalar $\lambda(\phi^2-1)^2$ theory}, Phys. Rev. D 44 (1991) 1147.

\bibitem{domerosh} P. Dorey, K. Mersh, T. Romanczukiewicz, Ya. Shnir, {\it Kink-antikink collisions in the $\phi^6$ model}, Phys. Rev. Lett. 107 (2011).

\bibitem{sgno} F.C. Simas, Adalto R. Gomes, K.Z. Nobrega, J.C.R.E. Oliveira, {\it Suppression of two-bounce windows in kink-antikink collisions}, JHEP 1609 (2016) 104. 

\bibitem{weig1} I. Takyi and H. Weigel, {\it Collective coordinates in one-dimensional soliton models revisited}, Phys. Rev. D 94, 085008 (2000).



\bibitem{dedekecrsa} A. Demirkaya, R. Decker, P. G. Kevrekidis, I. C. Christov, A. Saxena, {\it Kink Dynamics in a Parametric $\phi^6$ System: A Model With Controllably Many Internal Modes}, JHEP 12(2017)071.

\bibitem{gakuli1} V.A. Gani, A.E. Kudryavtsev and M.A. Lizunova, {\it Kink interactions in the (1+1)-dimensional $\phi^6$ model}, Phys. Rev. D 89 (2014) 125009.

\bibitem{wei} H. Weigel, {\it Kink-Antikink Scattering in $\phi^4$ and $\phi^6$ Models}, J. Phys. Conf. Ser. 482 (2014) 012045.

\bibitem{roman1} T. Romanczukiewicz, {\it Could the primordial radiation be responsible for vanishing of topological defects?}, Phys. Lett. B 773 (2017) 295.

\bibitem{belgani1} E. Belendryasova and Vakhid A. Gani, {\it Resonance phenomena in the $\phi^8$ kinks scattering}, J. Phys. Conf. Ser. 934 (2017), 012059.

\bibitem{ganilenliz1} V.A. Gani, V. Lensky and M.A. Lizunova, {\it Kink excitation spectra in the (1+1)-dimensional $\phi^8$ model}, JHEP 08 (2015) 147.

\bibitem{ekagani} Ekaterina Belendryasova and Vakhid A. Gani, {\it Scattering of the $\phi^8$ kinks with power-law asymptotics}, arXiv:1708.00403.

\bibitem{fv} Adalto R. Gomes, F.C. Simas, K.Z. Nobrega, P.P. Avelino, {\it False vacuum decay in kink scattering}, JHEP 1810 (2018) 192. 



\bibitem{halromshn} A. Halavanau, T. Romanczukiewicz, Ya. Shnir, {\it Resonance structures in coupled two-component $\phi^4$ model}, Phys. Rev. D86 (2012) 085027.

\bibitem{al1} A. Alonso Izquierdo, {\it Kink dynamics in a system of two coupled scalar fields in two space-time dimensions}, Physica D 365, 12 (2018). arXiv:1711.08784

\bibitem{al2}  A. Alonso Izquierdo, {\it Reflection, transmutation, annihilation, and resonance in two-component kink collisions}, Phys. Rev. D 97, 045016 (2018). arXiv:1711.10034

\bibitem{al3} A. Alonso Izquierdo, {\it Asymmetric kink scattering in a two-component scalar field theory model}, ArXiv:1901.03089.



\bibitem{peycam} M. Peyrard, D. K. Campbell, {\it Kink-antikink interactions in a modified sine-Gordon model}, Physica D Nonlinear Phenomena 9 (1983) 33-51.

\bibitem{gankud} V.A. Gani and A.E. Kudryavtsev, {\it Kink-antikink interactions in the double sine-Gordon equation and the problem of resonance frequencies}, Phys. Rev. E 60 (1999) 3305 [cond-mat/9809015].

\bibitem{sgn} F.C. Simas, A.R. Gomes, K.Z. Nobrega, {\it Degenerate vacua to vacuumless model and kink-antikink collisions}, Phys. Lett. B 775 (2017) 290.

\bibitem{gaaes} V. A. Gani, A. M. Marjaneh, A. Askari, E. Belendryasova and D. Saadatmand, {\it Scattering of the double sine-Gordon kinks}, Eur. Phys. J. C (2018) 78: 345.

\bibitem{bbv} D. Bazeia, E. Belendryasova, Vakhid A. Gani, {\it Scattering of kinks in a non-polynomial model}, J. Phys. Conf. Ser. 934 (2017) 012032.

\bibitem{bbv2} D. Bazeia, E. Belendryasova, Vakhid A. Gani, {\it Scattering of kinks of the sinh-deformed $\phi^4$ model}, Eur. Phys. J. C (2018) 78: 340.



\bibitem{magasaadmja} A. M. Marjaneh, V. A. Gani, D. Saadatmand, S. V. Dmitriev and K. Javidana, Multi-kink collisions in the $\phi^6$ model, JHEP 07 (2017) 028.

\bibitem{maassadm} A. M. Marjaneh, A. Askari, D. Saadatmand, S. V. Dmitriev, Extreme values of elastic strain and energy in sine-Gordon multi-kink collisions, Eur. Phys. J. B (2018) 91: 22.

\bibitem{sadmkev} D. Saadatmand, S.V. Dmitriev and P.G. Kevrekidis, High energy density in multisoliton collisions, Phys. Rev. D 92 (2015) 056005.

\bibitem{almasazhdi} Aliakbar Moradi Marjaneh, Danial Saadatmand, Kun Zhou, Sergey V. Dmitriev and Mohammad Ebrahim Zomorrodian, High energy density in the collision of N kinks in the $\phi^4$ model, Commun. Nonlinear Sci. Numer. Simul. 49 (2017) 30.

\bibitem{gan3} Vakhid A. Gani, Aliakbar Moradi Marjaneh, Danial Saadatmand, {\it Multi-kink scattering in the double sine-Gordon model},  arXiv:1901.07966.


\bibitem{gib} 	 John T. Giblin,  Lam Hui, Eugene A. Lim, I-Sheng Yang, {\it How to Run Through Walls: Dynamics of Bubble and Soliton Collisions}, Phys.Rev. D82 (2010) 045019.

\bibitem{graph1} T. Mashoff, M. Pratzer, V. Geringer, T. J. Echtermeyer,
M. C. Lemme, M. Liebmann, and M. Morgenstern, Nano Letters 10, 461 (2010).

\bibitem{graph2} R. D. Yamaletdinov, V. A. Slipko, Y. V. Pershin, {\it Kinks and antikinks of buckled graphene: A testing ground for the $\phi^4$ field model}, Phys. Rev. B 96, 094306 (2017).

\bibitem{ssh} W. P. Su, J. R. Schrieffer, A. J. Heeger, {\it Solitons in Polyacetylene},
Phys. Rev. Lett. 1979, 42, 1698-1701.

\bibitem{poly} L. Bernasconi, {\it Chaotic Soliton Dynamics in Photoexcited trans-Polyacetylene}, J. Phys. Chem. Lett.  6, 5, 908 (2015).









 


\bibitem{junker} G. Junker, Supersymmetric Methods in Quantum and Statistical Mechanics. Springer, 1996.

\bibitem{cookhasuk} F. Cooper, A. Khare, and U. Sukhatme, Supersymmetry in Quantum Mechanics. Work Scientic, 2001.




\bibitem{flosva} G. Flores-Hidalgo and N. F. Svaiter, {\it .	
Constructing bidimensional scalar field theory models from zero mode fluctuations},  Phys. Rev. D 66, 025031 (2002).

\bibitem{boryur} M. Bordag and A. Yurov, {\it 	
Spontaneous symmetry breaking and reflectionless scattering data},  Phys. Rev. D 67, 025003 (2003).

\bibitem{bazbem1} D. Bazeia and F.S. Bemfica, {\it 	
From Supersymmetric Quantum Mechanics to Scalar Field Theories}, Phys. Rev. D 95, 085008 (2017).

\bibitem{bazlos1} D. Bazeia and L. Losano, {\it 	
A novel connection between scalar field theories and quantum mechanics}, EPL 121, 10006 (2018).


\bibitem{bfl} D. Bazeia, D.A. Ferreira, Elisama E.M. Lima, L. Losano, {\it 	
Novel results for kinklike structures and their connections to quantum mechanics}, Annals Phys. 395 (2018) 275.

\bibitem{blm} D. Bazeia, L. Losano, and J.M.C. Malbouisson, {\it 
Deformed defects}, Phys. Rev. D 66, 101701 (2002).

\bibitem{abl} C.A. Almeida, D. Bazeia, L. Losano, and J.M.C. Malbouisson, {\it 	
New results for deformed defects}, Phys. Rev. D 69, 067702 (2004).



\bibitem{b} K. G. Zloshchastiev, {\it Coexistence of black holes and a long-range scalar field in cosmology}, Phys. Rev. Lett. 94 (2005) 121101.

\bibitem{whee} R. Ruffini and J. A. Wheeler, {\it Introducing the black hole}, Phys. Today 24, No. 1, 30
(1971).

\bibitem{nh} P. Bizon, {\it Gravitating solitons and hairy black holes}, Acta Phys. Pol. B 25, 877 (1994).

\bibitem{duf} M. J. Duff and J. T. Liu, {\it Anti-de Sitter black holes in gauged $N=8$ supergravity}, Nucl. Phys. B554, 237
(1999).

\bibitem{free} D. Z. Freedman, S. S. Gubser, K. Pilch, and N. P. Warner, {\it 	
Continuous distributions of D3-branes and gauged supergravity},  J.
High Energy Phys. 07 (2000) 038.

\bibitem{c} Qiang Wen, {\it Strategy to Construct Exact Solutions in Einstein-Scalar Gravities}, Phys. Rev. D 92, 104002 (2015)


\end{thebibliography}
\end{document}